\documentclass[12pt,oneside]{article}
\usepackage{amssymb,amsmath,latexsym,amsthm,amsfonts}
\usepackage[english]{babel}
\usepackage[T1]{fontenc}
\usepackage{color,hyperref}
\usepackage{multirow}
\newtheorem{prop}{Proposition}

\numberwithin{ex}{section}
\numberwithin{rem}{section}
\numberwithin{equation}{section}
\numberwithin{thm}{section}
\numberwithin{lem}{section}
\numberwithin{coro}{section}

\def\1g{1\hskip -3pt \mbox{l}}

\small\normalsize

\title{Testing normality for unconditionally heteroscedastic macroeconomic variables}

    \author{
{\sc
Hamdi Ra\"{i}ssi\footnote{Instituto de Estad\'{\i}stica, PUCV. Address:  Av. Errazuriz 2734, Valpara\'{\i}so, V regi\'{o}n, Chile. E-mail: hamdi.raissi@pucv.cl. This paper was supported by CONICYT-FONDECYT under grant  $N^{\circ} 1160527$.
The author gratefully acknowledge Steffen Gr{\o}nneberg and Genaro Sucarrat for helpful discussions during his stay at the BI Norwegian Business School.}}}

\begin{document}

\maketitle  \noindent {\em Abstract:} In this paper the testing of normality for unconditionally heteroscedastic macroeconomic time series is studied. It is underlined that the classical Jarque-Bera test (JB hereafter) for normality is inadequate in our framework. On the other hand it is found that the approach which consists in correcting the heteroscedasticity by kernel smoothing for testing normality is justified asymptotically. Nevertheless it appears from Monte Carlo experiments that such methodology can noticeably suffer from size distortion for samples that are typical for macroeconomic variables. As a consequence a parametric bootstrap methodology for correcting the problem is proposed. The innovations distribution of a set of inflation measures for the U.S., Korea and Australia are analyzed.



\section{Introduction}

In the econometric literature, the Jarque Bera (1980) test is routinely used to assess the normality of variables.
The properties of this test are well documented for stationary conditionally heteroscedastic processes. For instance Fiorentini, Sentana and Calzolari (2003), Lee, Park and Lee (2010) and Lee (2012) investigated the JB test in the context of GARCH models.
However few studies
are available on the distributional specification of time series in presence of unconditional heteroscedasticity. Drees and St\u{a}ric\u{a} (2002),
Mikosch and St\u{a}ric\u{a} (2004) and Fry\'{z}lewicz (2005) investigated the possibility of modelling financial returns by nonparametric methods. To this end, Drees and St\u{a}ric\u{a} (2002) and
Mikosch and St\u{a}ric\u{a} (2004) examined the distribution of S\&P500 returns corrected from heteroscedasticity. On the other hand
Fry\'{z}lewicz (2005)
pointed out that large sample kurtosis for financial time series may be explained by non constant unconditional variance.
In general we did not found references that specifically address the problem of assessing the distribution
of unconditionally heteroscedastic time series.
Note that non constant variance constitutes an important pattern for time series in general, and macroeconomic variables in particular. Reference can be made to Sensier and van Dijk (2004) who found that most of the 214 U.S. macroeconomic time series they studied have a time-varying variance.
In this paper we aim to provide a reliable methodology for testing normality for small samples time series with non constant unconditional variance.


The structure of the paper is as follows. In Section 2 we first set the dynamics
ruling the observed process. In particular the unconditional heteroscedastic structure of the errors is given. The inadequacy of the standard JB test in our framework is highlighted. The approach consisting in correcting the errors from the heteroscedasticity for building a JB test is presented. We then introduce a parametric bootstrap procedure that is intended to improve the normality testing for unconditionally heteroscedastic macroeconomic time series. In Section \ref{numerical} numerical experiments are conducted to
shed some light on the finite sample behaviors of the studied tests. In particular it is found that when estimating the non constant variance structure by kernel smoothing, a correct bandwidth choice is a necessary condition for the good implementation of the normality tests based on heteroscedasticity correction.
We illustrate our
outputs examining the distributional properties of the U.S., Korean and Australian GDP implicit price deflators.


\section{Testing normality in presence of unconditional heteroscedasticity}

We consider processes $(y_{t,n})$ which can be written as

\begin{eqnarray}
y_{t,n}&=&\omega_0+x_{t,n},\nonumber\\
x_{t,n}&=&\sum_{i=1}^pa_{0i}x_{t-i,n}+u_{t,n},\label{model}
\end{eqnarray}
where $y_{1,n},\dots,y_{n,n}$ are available, $n$ being the sample size and $E(x_{t,n})=0$. The conditional mean of $x_{t,n}$ is driven by the autoregressive parameters $\theta_0=(a_{01},\dots,a_{0p})'$. We make the following
assumption on the conditional mean.\\

\textbf{Assumption A0:}\: The ${a}_{0i}\in\mathbb{R}$, $1\leq i\leq p$, are
    such that $\det ({a}(z))\neq 0$ for all $|z|\leq 1$, with
    ${a}(z)=1-\sum_{i=1}^{p}{a}_{0i} z^i$.\\

In the assumption {\bf A1} below, the well known rescaling device introduced by
Dahlhaus (1997) is used to specify the errors process $(u_{t,n})$. For a random
variable $v$ we define $\|v\|_q=(E|v|^q)^{1/q}$, with $E|v|^q<1$ and $q\geq1$.\\

\textbf{Assumption A1:}\: We assume that $u_{t,n}=h_{t,n}\epsilon_t$ where:
\begin{itemize}
\item[(i)] $h_{t,n}\geq c>0$ for some constant $c>0$, and satisfies $h_{t,n}=g(t/n)$, where $g(r)$ is a measurable deterministic function on the interval $(0,1]$, such that $\sup_{r\in(0,1]}|g(r)|<\infty$. The function $g(.)$ satisfies a Lipschitz condition piecewise on a finite number of some sub-intervals that partition $(0,1]$.
\item[(ii)] The process $(\epsilon_t)$ is iid and such that $E(\epsilon_t)=0,$ $E(\epsilon_t^2)=1$, and $(E(\|\epsilon_t\|^{8\nu})<\infty$ for some $\nu>1$.\\
\end{itemize}

The non constant variance induced by {\bf A1(i)} allows for a wide range of non stationarity patterns commonly faced in practice, as for instance abrupt shifts, smooth changes or cyclical behaviors. Note that
in the zero mean AR case, tools needed to carry out the Box and Jenkins specification-estimation-validation modeling cycle,
are provided in Patilea and Ra\"{\i}ssi (2013) and Ra\"{\i}ssi (2015). For $\omega_0\neq0$ define the estimator
$\hat{\omega}=n^{-1}\sum_{t=1}^{n}y_{t,n}$, and $x_{t,n}(\omega)=y_{t,n}-\omega$ for any $\omega\in\mathbb{R}$. Writing $\hat{\omega}-\omega_0=n^{-1}\sum_{t=1}^{n}x_{t,n}$, it can be shown that
\begin{equation}\label{const}
\sqrt{n}(\hat{\omega}-\omega_0)=O_p(1),
\end{equation}
using the Beveridge-Nelson decomposition.
Now let
\begin{equation}\label{condmean}
\hat{\theta}(\omega)=\left(\Sigma_{\underline{x}}(\omega)\right)^{-1}\Sigma_x(\omega),
\end{equation}
where $$\Sigma_{\underline{x}}(\omega)=
n^{-1}\sum_{t=1}^{n}\underline{x}_{t-1,n}(\omega)\underline{x}_{t-1,n}(\omega)'\quad\mbox{and}\quad
\Sigma_x(\omega)=n^{-1}\sum_{t=1}^{n}\underline{x}_{t-1,n}(\omega)x_{t-1,n}(\omega),$$ with $\underline{x}_{t-1,n}(\omega)=(x_{t-1,n}(\omega),\dots,x_{t-p,n}(\omega))'$. With these notations define the OLS
estimator $\hat{\theta}(\hat{\omega})$ and the unfeasible estimator $\hat{\theta}(\omega_0)$. Straightforward computations give
$\sqrt{n}(\hat{\theta}(\hat{\omega})-\hat{\theta}(\omega_0))$ $=o_p(1)$, so that using the results of Patilea and Ra\"{\i}ssi (2012) we have
\begin{equation}\label{autoreg}
\sqrt{n}(\hat{\theta}(\hat{\omega})-\theta_0)=O_p(1).
\end{equation}

Once the conditional mean is filtered in accordance to (\ref{const}) and (\ref{autoreg}), we can proceed to
the test of the following hypotheses:
$$H_0:\:\epsilon_t\sim\mathcal{N}(0,1)\quad\mbox{vs.}\quad H_1:\epsilon_t\:\mbox{ has a different distribution},$$
with the usual slight abuse of interpretation inherent of the use JB test for normality testing.
Clearly the skewness and kurtosis of $u_{t,n}$
correspond to those of $\epsilon_t$. However in practice $E(u_{t,n}^3)=0$ and $E(u_{t,n}^4)=3$ is checked
using the JB test statistic:

\begin{equation}\label{jb}
  Q_{JB}^u=n\left[Q_{JB}^{S,u}+Q_{JB}^{K,u}\right],
\end{equation}
where

$$Q_{JB}^{S,u}=\frac{\hat{\mu}_3^2}{6\hat{\mu}_2^3}\quad\mbox{and}\quad Q_{JB}^{K,u}=\frac{1}{24}\left(\frac{\hat{\mu}_4}{\hat{\mu}_2^2}-3\right)^2,$$
with $\hat{\mu}_j=n^{-1}\sum_{t=1}^{n}(\hat{u}_{t,n}-\bar{\hat{u}})^j$ and $\bar{\hat{u}}=n^{-1}\sum_{t=1}^{n}\hat{u}_{t,n}$. The $\hat{u}_{t,n}$'s are the residuals obtained from
the estimation step. Let us denote by $\Rightarrow$
convergence in distribution. If we suppose the process $(u_t)$
homoscedastic ($g(.)$ is constant), then the standard result $Q_{JB}^u\Rightarrow\chi_2^2$ is retrieved (see Yu (2007), Section 2.2). However under {\bf A0} and
{\bf A1} with $g(.)$ non constant (the unconditionally heteroscedastic case)
we have:

%

\begin{equation}\label{divergence2}
Q_{JB}^{K,u}=\frac{1}{24}\left[\kappa_2\left(E(\epsilon_t^4))-3\right)
+3\left(\kappa_2-1\right)\right]+o_p(1),
\end{equation}
where $\kappa_2=\frac{\int_{0}^{1}g^4(r)dr}{\left(\int_{0}^{1}g^2(r)dr\right)^2}$.
Hence if we suppose the errors process unconditionally
heteroscedastic with $E(\epsilon_t^4)=3$, we have $Q_{JB}^u=Cn+o_p(n)$ for some strictly positive
constant $C$. As a consequence, the classical JB test will tend to detect spuriously departures from the null hypothesis of a normal distribution in our framework. This argument was used by Fry\'{z}lewicz (2004) to underline that unconditionally heteroscedastic specifications can cover financial time series that typically exhibit an excess of kurtosis.\\

In order to assess the distribution of S\&P500 returns, Drees and St\u{a}ric\u{a} (2002)
considered data corrected from heteroscedasticity, using a kernel estimator of the
variance. We will follow this approach in the sequel considering

$$\hat{h}_{t,n}^2=\sum_{i=1}^nw_{ti}(\hat{u}_{i,n}-\bar{\hat{u}})^2,\qquad 1\leq t\leq n,$$
with $w_{ti}=\left(\sum_{j=1}^nK_{tj}\right)^{-1}K_{ti}$,
$K_{ti}= K((t-i)/nb)$ if $t\neq i$ and $K_{ii}=0,$
where $K(\cdot)$ is a kernel function on the real line and $b$ is the bandwidth. The following assumption is needed for our variance estimator.\\

\textbf{Assumption A2:} \, (i) The kernel $K(\cdot)$ is a bounded density function defined on the real line such that $K(\cdot)$ is nondecreasing on $(-\infty, 0]$ and decreasing on $[0,\infty)$ and $\int_\mathbb{R} v^2K(v)dv < \infty$. The function $K(\cdot)$ is differentiable except a finite number of points and the derivative $K^\prime(\cdot)$  satisfies $\int_{\mathbb{R}}|x K^\prime (x)| dx < \infty.$
    Moreover, the Fourier Transform $\mathcal{F}[K](\cdot)$ of $K(\cdot)$ satisfies $\int_{\mathbb{R}}  \left| s\right|^\tau \left| \mathcal{F}[K](s) \right|ds <\infty$ for some $\tau>0$.

    (ii) The bandwidth $b$ is taken in the range $\mathfrak{B}_n = [c_{min} b_n, c_{max} b_n]$ with $0< c_{min}< c_{max}< \infty$ and $nb_n^{4-\gamma} + 1/nb_n^{2+\gamma} \rightarrow 0$ as $n\rightarrow \infty$, for some small $\gamma >0$.\\

Let $\hat{\epsilon}_t=(\hat{u}_{t,n}-\bar{\hat{u}})/\hat{h}_{t,n}$. We are now ready to consider the following JB test statistic:

\begin{equation*}\label{jb}
  Q_{JB}^{\epsilon}=n\left[Q_{JB}^{S,\epsilon}+Q_{JB}^{K,\epsilon}\right],
\end{equation*}
where

$$Q_{JB}^{S,\epsilon}=\frac{\hat{\nu}_3^2}{6\hat{\nu}_2^3}\quad\mbox{and}\quad Q_{JB}^{K,\epsilon}=\frac{1}{24}\left(\frac{\hat{\nu}_4}{\hat{\nu}_2^2}-3\right)^2,$$
with $\hat{\nu}_j=n^{-1}\sum_{t=1}^{n}\hat{\epsilon}_t^j$. The following proposition gives the asymptotic distribution of $Q_{JB}^{\epsilon}$. 

\begin{prop}\label{propostu}
Under the assumptions {\bf A0}, {\bf A1} and {\bf A2}, we have as $n\to\infty$

\begin{equation}\label{res}
Q_{JB}^{\epsilon}\Rightarrow\chi_2^2,
\end{equation}
uniformly with respect to $b\in\mathfrak{B}_n$.
\end{prop}

Proposition 1 can be proved using the same arguments given in Yu (2007), together with those of the proof of Proposition 4 in Patilea and Ra\"{\i}ssi (2014). Therefore we skip the proof.
For building a test using the above result, we will consider the normal kernel in the next section. On the other hand we suggest to choose a bandwidth by minimizing the cross-validation (CV) criterion (see Wasserman (2006,p69-70)), unless otherwise specified.
The test obtained using (\ref{res}) and the above settings will be denoted by $T_{cv}$. The standard test, that does not take into account the unconditional heteroscedasticity, is denoted by $T_{st}$.

For high frequency time series it is reasonable to suppose that the approximation (\ref{res}) is satisfactory when the bandwidth is carefully chosen. Nevertheless considering the above sophisticated procedure for small $n$ is questionable. Therefore we propose to apply the following parametric bootstrap algorithm inspired from Francq and Zako\"{\i}an (2010,p335).\\

\begin{itemize}
  \item[1-] Generate $\epsilon_t^{(b)}\sim\mathcal{N}(0,1)$, $1\leq t\leq n$, build the bootstrap errors $u_{t,n}^{(b)}=\epsilon_t^{(b)}\hat{h}_{t,n}$, and the bootstrap series $y_t^{(b)}$ using (\ref{model}), but with $\hat{\omega}$ and $\hat{\theta}(\hat{\omega})$ (see (\ref{const}) and (\ref{condmean})).
  \item[2-] Estimate the autoregressive parameters and a constant as in (\ref{model}), but using the $y_t^{(b)}$'s. Build the kernel estimators $\hat{h}_{t,n}^{(b)}$ from the resulting residuals $\hat{u}_{t,n}^{(b)}$.
  \item[3-] Compute $\hat{\epsilon}_{t,n}^{(b)}=\hat{u}_{t,n}^{(b)}/\hat{h}_{t,n}^{(b)}$ for $t=1,\dots,n$. Compute $Q_{JB}^{\epsilon,(b)}$.
  \item[4-] Repeat the steps 1 to 3 B times for some large B. Use the $Q_{JB}^{\epsilon,(b)}$'s to compute the p-values of the bootstrap JB test.
\end{itemize}

The test obtained using the above parametric bootstrap procedure is denoted by $T_{boot}$.

\section{Numerical illustrations}
\label{numerical}

The finite sample properties of the $T_{st}$, $T_{cv}$ and $T_{boot}$ tests are first examined by means of Monte Carlo experiments. The distribution of
the U.S., Korean and Australian GDP implicit price deflator is then investigated.
Throughout this section the asymptotic nominal level of the tests is 5\%. In the sequel, we fixed $B=499$.

\subsection{Monte Carlo experiments}
\label{mce}

We simulate $N=1000$ trajectories of AR(1) processes:

\begin{equation}\label{ar1}
y_{t,n}=a_0y_{t-1,n}+u_{t,n},
\end{equation}
where $a_0=0.4$ and $u_{t,n}=h_{t,n}\epsilon_t$ with $\epsilon_t$ iid(0,1). Under the null hypothesis we set $\epsilon_t\sim\mathcal{N}(0,1)$. On the other hand under the alternative hypothesis $\epsilon_t=\cos(\delta)v_t+\sin(\delta)w_t$ is taken, with $v_t\sim\mathcal{N}(0,1)$, $(\sqrt{2}w_t+1)\sim\chi_1^2$, $0<\delta\leq\frac{\pi}{2}$, $v_t$ and $w_t$ being mutually independent.
In order to study the case where the series are actually homoscedastic, we set
$h_{t,n}=1$. For the heteroscedastic case, the variance structure is given by

\begin{equation}\label{simhet}
h_{t,n}=1+2\exp\left(t/n\right)+0.3(1+t/n)\sin\left(5\pi t/n+\pi/6\right).
\end{equation}
In such situation the variance structure exhibits a global monotone behavior together with a cyclical/seasonal component that is common in macroeconomic data (see e.g. Trimbur, and Bell (2010) for seasonal effects in the variance). In all our experiments, the mean in (\ref{ar1}) is treated as unknown. More precisely the AR parameter in (\ref{ar1}) is estimated using $y_{t,n}-\hat{\omega}$, where $\hat{\omega}$ is given in (\ref{const}), and then the resulting centered residuals are used to compute the test statistics.\\

The outputs obtained under the null hypothesis are first analyzed. The results are given
in Table \ref{tab1} for the homoscedastic case and in Table \ref{tab2} for the heteroscedastic case. Noting that macroeconomic time series with noticeable heteroscedasticity are relatively large but smaller than $n=400$ in general, a special emphasis will be put on interpreting results for samples $n=100,200,400$. Since $N=1000$ processes are simulated, and under the hypothesis that the finite sample size of a given test
is $5\%$, the relative rejection frequencies should be between the significant
limits 3.65\% and 6.35\% with probability 0.95. The outputs outside these confidence bands will be displayed in bold type.\\

From Table \ref{tab1}, it appears that the
$T_{cv}$ is oversized for small samples ($n=100$ and $n=200$). This could be explained by the fact that this test is too much sophisticated for the standard case. When the samples are increased the relative rejection frequencies become close to the 5\% ($n=400$ and $n=800$). On the other hand the $T_{st}$ and $T_{boot}$ tests have good results for all the samples. Of course if there is no evidence of heteroscedasticity, the simple $T_{st}$ should be used. However Table \ref{tab1} reveals that in case of doubt,
the use of the $T_{boot}$ is a good alternative.

In the heteroscedastic case, it is seen from Table \ref{tab2} that the $T_{st}$ test fails to control the type I error as $n\to\infty$. This was expected from (\ref{divergence2}). Next it seems that the relative rejection frequencies of the $T_{cv}$ test are somewhat far from the nominal level 5\%, even when $n=800$. From Table \ref{tab2} it also emerges that the $T_{boot}$ control reasonably well the type I error. Therefore
we can draw the conclusion that the $T_{boot}$ gives a substantial improvement
for samples that are typical for heteroscedastic macroeconomic variables.

Note that the $T_{cv}$ test have better results for larger samples ($n\gg1000$). For instance conducting similar experiences to those of Table \ref{tab2}, we obtained 7.4\% rejections for $n=1600$ and 6.9\% rejections for $n=3200$. Hence the potential improvements of the $T_{boot}$ in comparison to the $T_{cv}$ should become slight as $n\to\infty$. For this reason if high frequency time series are analyzed, the $T_{cv}$ should certainly be preferred to the computationally intensive $T_{boot}$.

In general it is important to point out that the bandwidth must be carefully selected to ensure a good implementation of the $T_{boot}$ and $T_{cv}$ tests. Its turns out from our experiments that selecting the bandwidth by cross-validation leads to  relatively correct results. Indeed we found that the rejection frequencies of the
$T_{cv}$ converge to the 5\%, and that the rejection frequencies of the $T_{boot}$
remain close to the nominal level in such a case. However other choices can deteriorate the control of the type I errors.
For instance let us consider the $T_{f}$ test which consists in correcting the heteroscedasticity, but with fixed bandwidth as $\gamma(\hat{\sigma}^2/n)^{0.2}$, where $\hat{\sigma}^2$ is the sample variance and $\gamma$ is a constant. The corresponding bootstrap test will be denoted by $T_{f,boot}$. Here the normal kernel is kept. We only study the heteroscedastic case. The results given in Table \ref{tabgam} show that the rejection frequencies are strongly affected by this way of selecting the bandwidth. 

Finally let us point out that when the heteroscedastic structure is relatively easy to estimate (for instance if the sinus part is removed in (\ref{simhet})), we found better results (not displayed here) for the $T_{boot}$ and $T_{cv}$ tests in comparison to those of Table \ref{tab2}.\\

\begin{figure}[h]\!\!\!\!\!\!\!\!\!\!
\vspace*{6.8cm}

\vspace*{-0.9 cm}

\protect \includegraphics{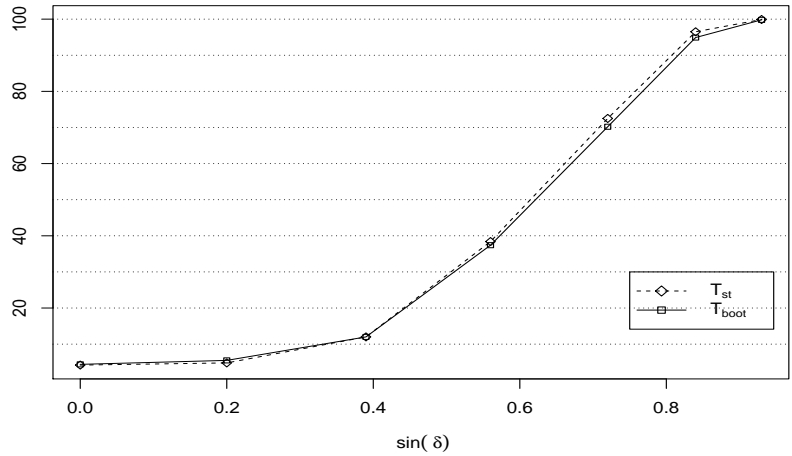} \caption{\label{powerfig}
{\footnotesize Empirical power (in \%) of the $T_{st}$ and $T_{boot}$ tests in the homoscedastic case.}}
\end{figure}

Now we turn to the analysis of the behavior of the tests under the alternative hypothesis. For a fair comparison we only studied the $T_{st}$ and $T_{boot}$ in the homoscedastic case. The sample size $n=100$ is fixed and recall that the parameter $\delta$ defines the departures from the null hypothesis. The outputs of our simulations, displayed in Figure \ref{powerfig}, show that the $T_{boot}$ test does not suffer from a lack of power in comparison to the $T_{st}$. In conclusion it turns out that the $T_{boot}$ improves the distribution analysis, in the sense that it ensures a good control of the type I error, but without entailing noticeable loss of power.

\subsection{Real data analysis}
\label{rda}

The inflation measures data are commonly used to analyze macroeconomic facts. Reference can be made to the numerous empirical papers studying the relation between price levels and money supply (see e.g. Jones and Uri (1986)). On the other hand inflation is of great importance in finance, as many central banks adjust their interest rates in view of targeting a certain
inflation level. Accordingly, constructing valid confidence intervals for inflation forecasts may be often crucial. In such kind of investigations clearly the distributional analysis can help to build a model for the data. In a stationary setting, authors aimed to detected ARCH effects assessing asymmetry and/or leptokurticity in inflation variables following Engle (1982) (see Broto and Ruiz (2008p22) among others). In the same way, it is reasonable to think that a test for normality taking into account the time-varying variance, can help to choose between a deterministic specification, as in {\bf A1}, and the case where in addition to unconditional heteroscedasiticity, second order dynamics are present (as in the case of spline-GARCH processes introduced by Engle and Rangel (2008)). In other words, once the unconditional heteroscedasiticity is removed from $u_t=h_t\epsilon_t$, the JB tests can help to decide whether ARCH effects are present or not in $(\epsilon_t)$.

\begin{figure}[h]\!\!\!\!\!\!\!\!\!\!
\vspace*{12.3cm} 


\protect \includegraphics{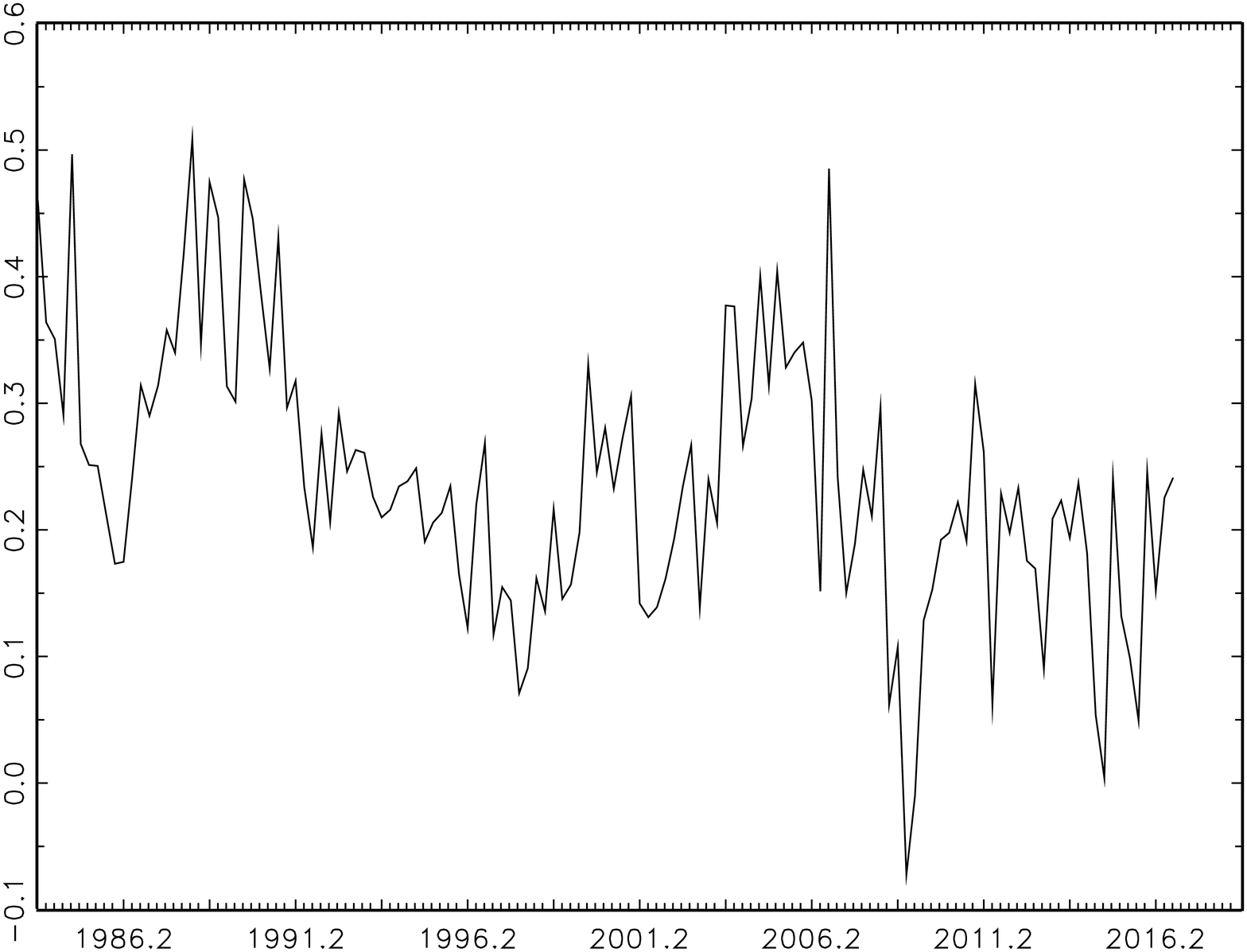} \protect \includegraphics{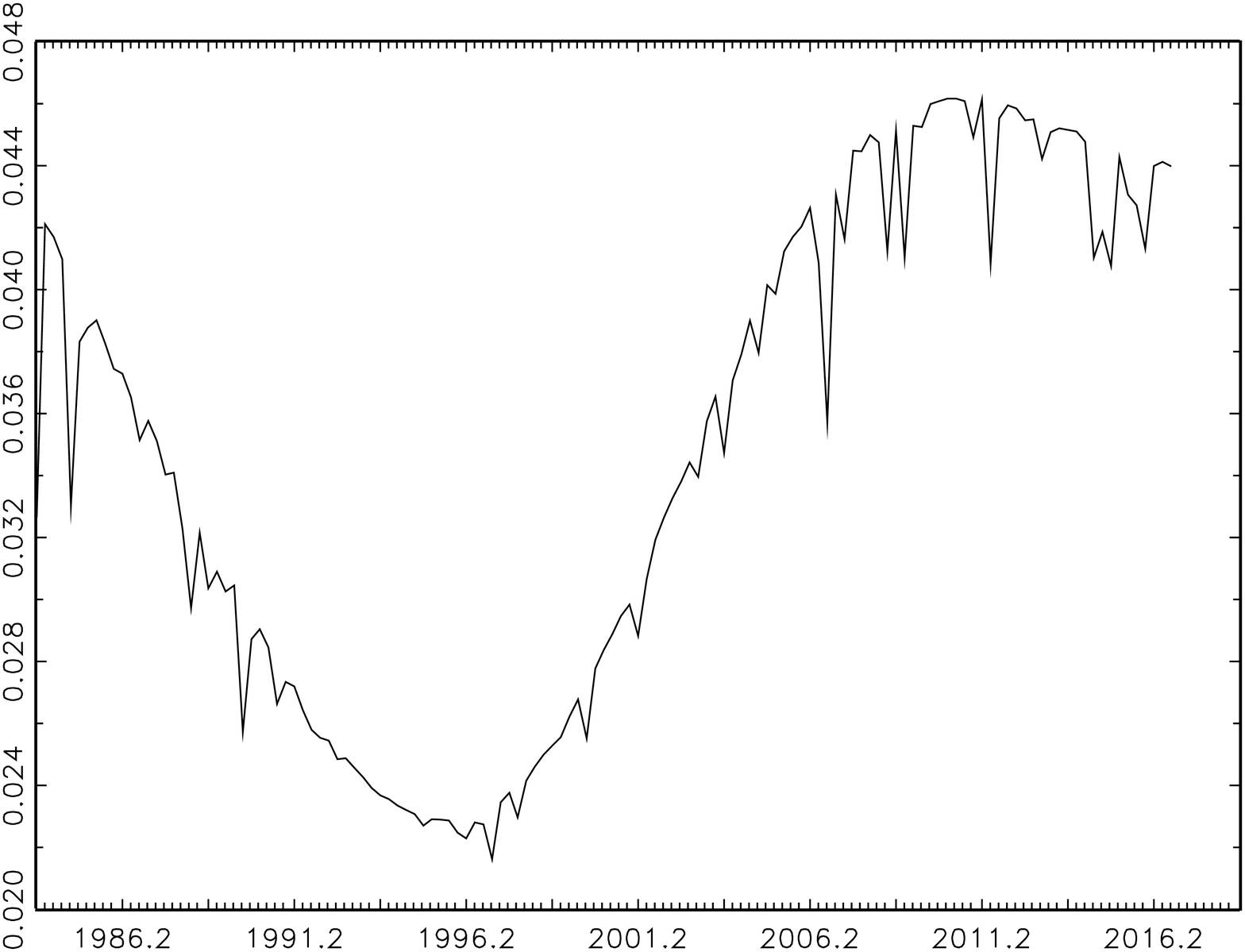}
\protect \includegraphics{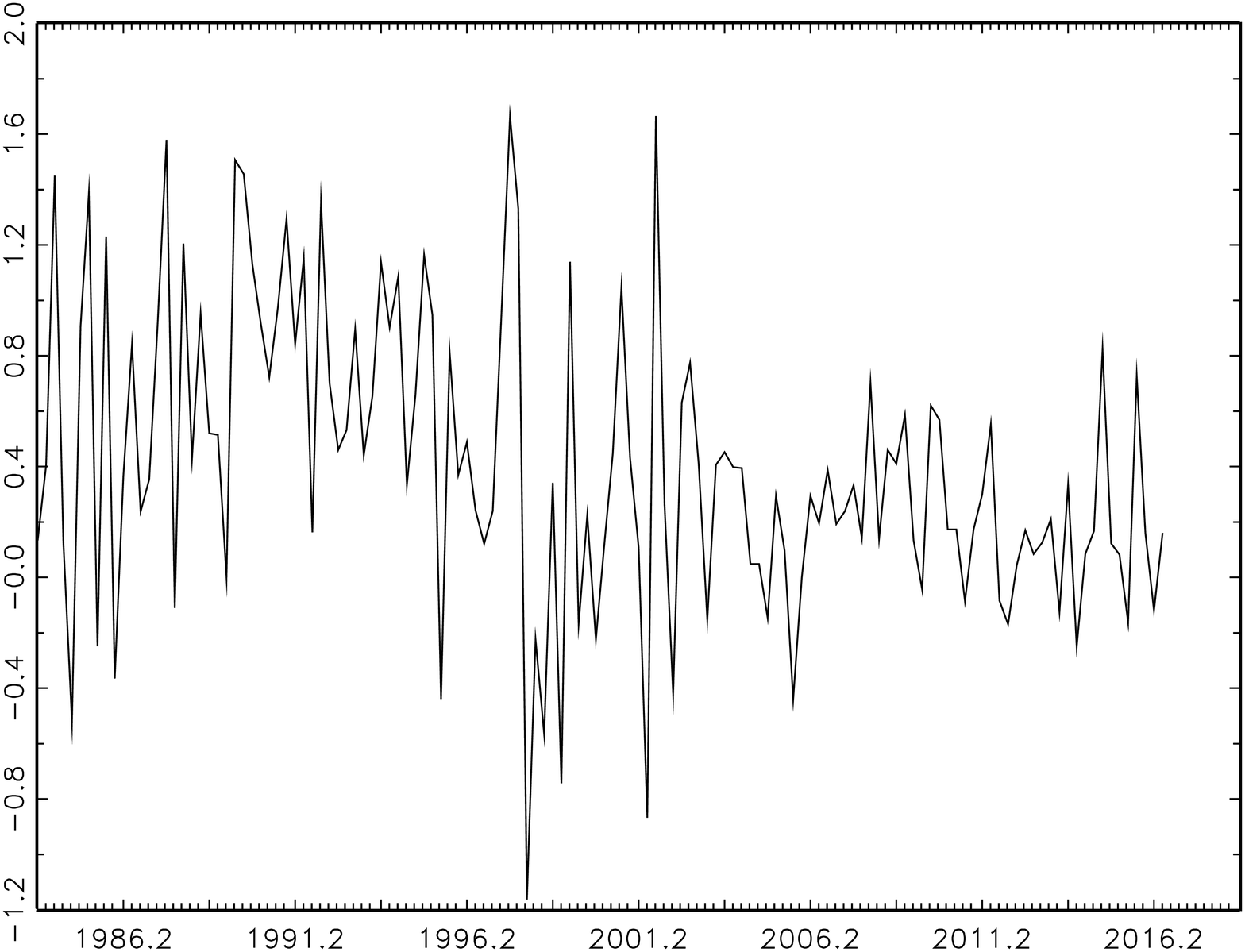} \protect \includegraphics{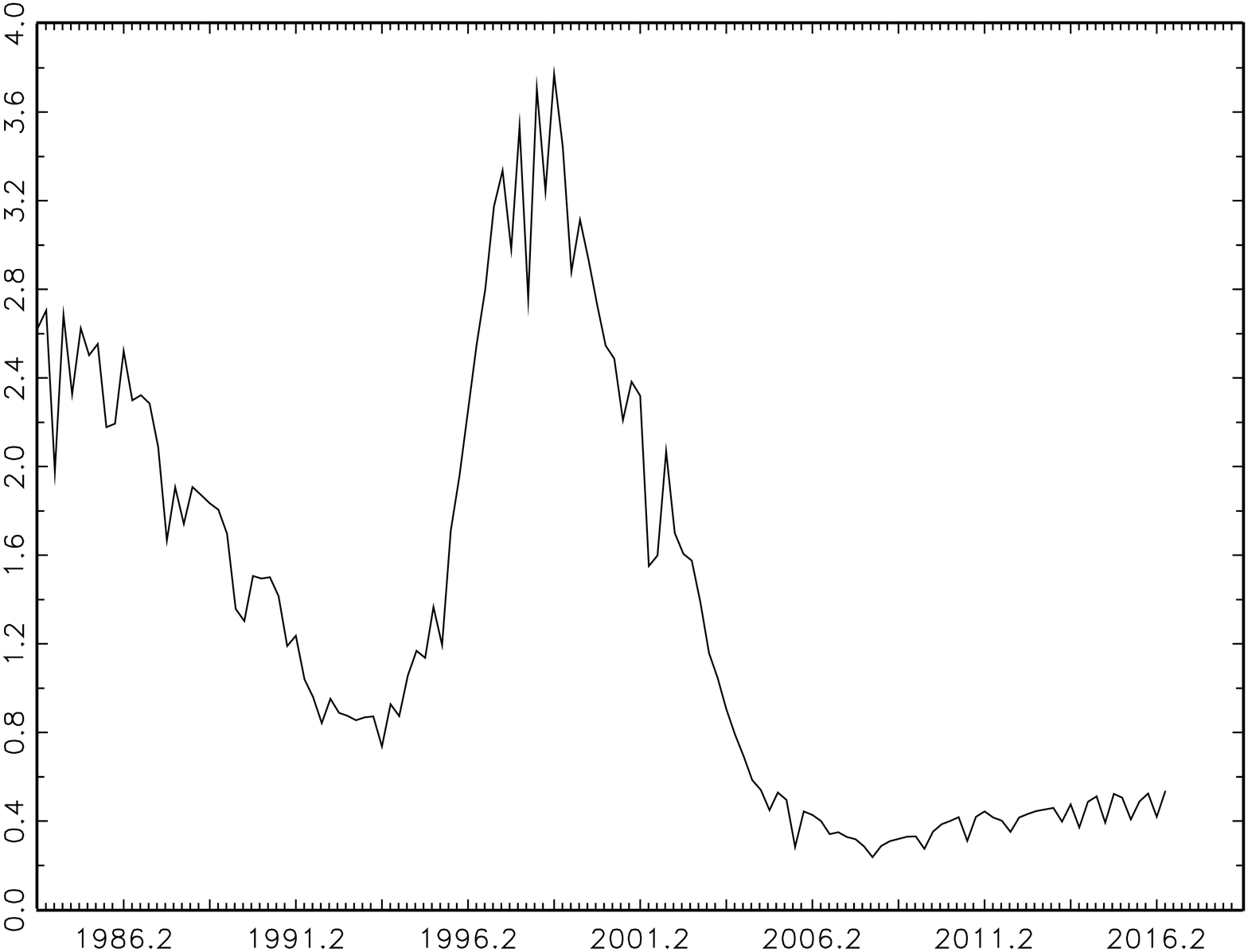}
\protect \includegraphics{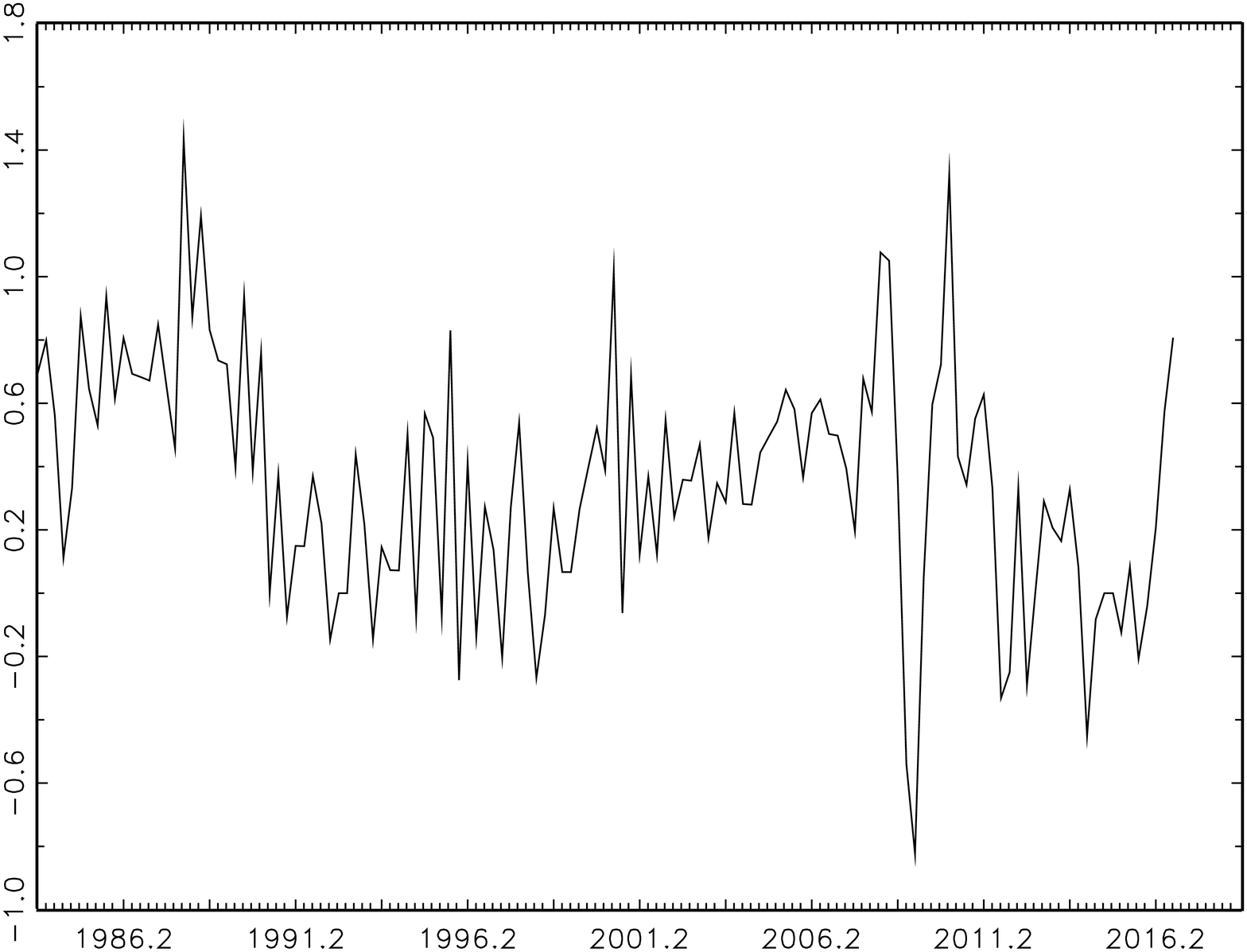} \protect \includegraphics{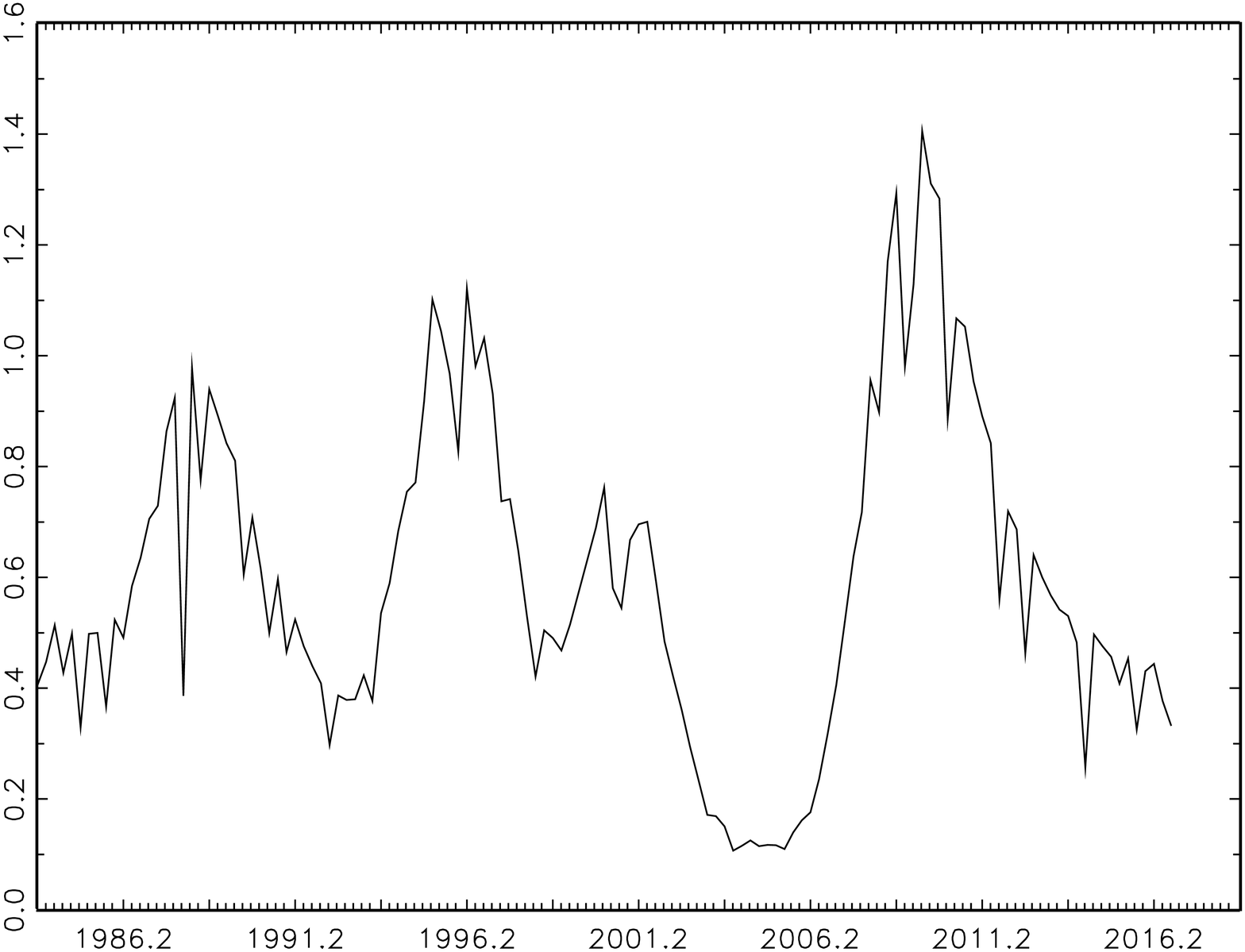}
\caption{\label{data}
{\footnotesize The log differences of the quarterly U.S. (top left panel), Korean (middle left panel) and Australian (bottom left panel) GDP implicit price deflators from 10/01/1983 to 01/01/2017 ($n=132$). The corresponding estimations of the innovations variance are on the right. Data source: The research division of the federal reserve bank of Saint Louis, fred.stlouisfed.org.}}
\end{figure}

In this part we will study the normality of the log differences of the quarterly GDP implicit price deflators for the U.S., Korea and Australia from 10/01/1983 to 01/01/2017 ($n=132$). More precisely we use $y_{t,n}=100\log\left(GDP_{t,n}/GDP_{t-1,n}\right)$. The data can be downloaded from the webpage of the research division of the federal reserve bank of Saint Louis: https://fred.stlouisfed.org. The studied variables plotted in Figure \ref{data} seem to show cyclical heteroscedasticity. In the case of Korea we can suspect a global decreasing behavior leading to a stabilization after the Asian crisis.
The times series are first filtered according to (\ref{model}). The non correlation of the residuals is tested using the adaptive portmanteau test of Patilea and Ra\"{\i}ssi (2013). On the other hand we applied tests for second order dynamics developed by
Patilea and Ra\"{\i}ssi (2014). The outputs (not displayed here) show that the hypothesis of no ARCH effects cannot be rejected. Hence the deterministic specification of the time-varying variance in {\bf A1} seems valid.
Once the linear dynamics of the series seem captured in an appropriate way, the tests considered in this paper are applied to the residuals. The results are given in Table \ref{tab3}. When the null hypothesis of normality is rejected at the 5\% level, the p-value is displayed in bold type. It emerges that the outputs of the $T_{boot}$ test are in general clearly different from those of the $T_{cv}$ and $T_{st}$ tests. The p-values of the $T_{cv}$ are all lower than those of the $T_{boot}$. Note that in the case of the U.S. GDP implicit price deflator, the difference between the $T_{boot}$ on one hand, and the $T_{st}$, $T_{cv}$ tests on the other hand, lead to different conclusions. In view of the outputs obtained from the simulations experiments, it is reasonable to decide that the normality assumption cannot be rejected for the U.S. data. It is likely that
rejecting normality will suggest more sophisticated models, and could entail misspecifications for the confidence intervals of the forecasts by fitting a heavy tailed distribution to the U.S. data.

\section*{References}
\begin{description}
\item[]{\sc Broto, C., and Ruiz, E.} (2008) Testing for conditional heteroscedasticity in the components of inflation. Working document, banco de Espa\~{n}a.
\item[]{\sc Dahlhaus, R.} (1997) Fitting time series models to nonstationary processes. \textit{Annals of Statistics} 25, 1-37.
\item[]{\sc Drees, H., and St\u{a}ric\u{a}, C.} (2002) A simple non stationary model for stock returns. Preprint. Universit\"{a}t des Saaland.
\item[]{\sc Engle, R.F.} (1982) Autoregressive conditional heteroscedasticity with estimates of the variance of United Kingdom inflation. \textit{Econometrica} 50, 987-1007.
\item[]{\sc Engle, R.F., and Rangel, J.G.} (2008) The spline GARCH model for unconditional volatility and its global macroeconomic causes. \textit{Review of Financial Studies} 21, 1187-1222.
\item[]{\sc Fiorentini, G., Sentana, E., Calzolari, G.} (2004) On the validity of the Jarque-Bera normality test in conditionally heteroscedastic dynamic regression models. \textit{Economic Letters} 83, 307-312.
\item[] {\sc Francq, C., and Zako\"{i}an, J-M.} (2010) \textit{GARCH models : structure, statistical inference, and financial applications.}
    Wiley, Chichester.
\item[] {\sc Fry\'{z}lewicz, P.} (2005) Modelling and forecasting financial log-returns as locally stationary wavelet processes. \textit{Journal of Applied Statistics} 32, 503-528.
\item[] {\sc Jarque, C.M., and Bera, A.K.} (1980) Efficient tests for normality, homoscedasticity and serial independence of regression residuals. \emph{Economics Letters} 6, 255-259.
\item[]{\sc Jones, J.D., and Uri, N.} (1986) Money, inflation and causality (another look at the empirical evidence for the USA, 1953-84). \textit{Applied Economics} 19, 619-634.
\item[]{\sc Lee, S., Park, S., and Lee, T.} (2010) A note on the Jarque-Bera normality test for GARCH innovations. \textit{Journal of the Korean Statistical Society} 39, 93-102.
\item[]{\sc Lee, T.} (2010) A note on Jarque-Bera normality test for ARMA-GARCH innovations. \textit{Journal of the Korean Statistical Society} 41, 37-48.
\item[]{\sc Mikosch, T., and St\u{a}ric\u{a}, C.} (2004) Stock market risk-return inference. An unconditional non-parametric approach. Research report, the Danish national research foundation: Network in Mathematical Physics and Stochastics.
\item[] {\sc Patilea, V., and Ra\"{i}ssi, H.} (2012) Adaptive estimation of vector autoregressive models with time-varying variance: application to testing linear causality in mean. \textit{Journal of Statistical Planning and Inference} 142, 2891-2912.
\item[] {\sc Patilea, V., and Ra\"{i}ssi, H.} (2013) Corrected portmanteau tests for VAR models with time-varying variance. \textit{Journal of Multivariate Analysis} 116, 190-207.
\item[] {\sc Patilea, V., and Ra\"{i}ssi, H.} (2014) Testing second order dynamics for autoregressive processes in presence of time-varying variance. \emph{Journal of the American Statistical Association} 109, 1099-1111.
\item[]{\sc Sensier, M., and van Dijk, D.} (2004) Testing for volatility changes in U.S. macroeconomic time series. \textit{Review of Economics and Statistics} 86, 833-839.
\item[]{\sc Trimbur, T.M., and Bell, W.R.} (2010) Seasonal heteroscedasticity in time series data: modeling, estimation, and testing. In W. Bell, S. Holan, and T. Mc Elroy (Eds.), \emph{Economic Time Series: Modelling and Seasonality}. Chapman and Hall, New York.
\item[]{\sc  Wasserman, L.} (2006) \textit{All of Nonparametric Statistics}.
    Springer, New-York.
\item[]{\sc  Ra\"{i}ssi, H.} (2015) Autoregressive order identification for VAR models with non-constant variance. \textit{Communications in Statistics: Theory and Methods} 44, 2059-2078.
\item[]{\sc  Yu, H.} (2007) High moment partial sum processes of residuals in ARMA models and their applications. \textit{Journal of Time Series Analysis} 28, 72-91.
\end{description}

\newpage
\section*{Tables and Figures}

\begin{table}[hh]\!\!\!\!\!\!\!\!\!\!
\begin{center}
\caption{\small{Empirical size (in \%) of the studied tests for normality. The homoscedastic case.}}
\begin{tabular}{|c|c|c|c|c|}
 \hline
  $n$ & 100 & 200 & 400 & 800  \\
  \hline
  $T_{st}$ & 4.0 & 5.2 & 4.9 & 4.2  \\
  \hline
  $T_{cv}$ & {\bf 7.2} & {\bf 7.5} & 5.6 & 5.0  \\
  \hline
  $T_{boot}$ & 4.5 & 5.6 & 5.0 & 4.7  \\
  \hline
\end{tabular}
\label{tab1}
\end{center}
\end{table}

\begin{table}[hh]\!\!\!\!\!\!\!\!\!\!
\begin{center}
\caption{\small{Empirical size (in \%) of the studied tests for normality. The heteroscedastic case.}}
\begin{tabular}{|c|c|c|c|c|}
 \hline
  $n$ & 100 & 200 & 400 & 800  \\
  \hline
  $T_{st}$ & {\bf 8.7} & {\bf 13.0} & {\bf 11.5} & {\bf 19.1}  \\
  \hline
  $T_{cv}$ & {\bf 9.4} & {\bf 9.2} & {\bf 8.3} & {\bf 7.8}  \\
  \hline
  $T_{boot}$ & 4.4 & {\bf 6.5} & 6.3 & 6.3  \\
  \hline
\end{tabular}
\label{tab2}
\end{center}
\end{table}

\begin{table}[hh]\!\!\!\!\!\!\!\!\!\!
\begin{center}
\caption{\small{Empirical size (in \%) of the $T_{cv}$ and $T_{boot}$ tests for normality with fixed bandwidth. The heteroscedastic case.}}
\begin{tabular}{|c|c|c||c|c|}
 \hline
 $\gamma$ &\multicolumn{2}{c||}{$1$}&\multicolumn{2}{c|}{$1.5$}\\
 \hline
  $n$ & 100 & 200 & 100 & 200  \\
  \hline
  $T_{f}$ & {\bf 11.3} & {\bf 14.0} & {\bf 12.0} & {\bf 14.3}  \\
  \hline
  $T_{f,boot}$ & {\bf 6.8} & {\bf 10.2} & {\bf 7.6} & {\bf 11.1}  \\
  \hline
\end{tabular}
\label{tabgam}
\end{center}
\end{table}

\begin{table}[hh]\!\!\!\!\!\!\!\!\!\!
\begin{center}
\caption{\small{The p-values (in \%) of the tests for normality for GDP implicit price deflators for the U.S., Korea and Australia.}}
\begin{tabular}{c|c|c|c|}\cline{2-4}
    & \mbox{U.S.}  & \mbox{Korea} & \mbox{Australia} \\
  \hline
  \multicolumn{1}{|c|}{$T_{st}$} & {\bf 3.8} & 16.4 &  50.9  \\
  \hline
  \multicolumn{1}{|c|}{$T_{cv}$} & {\bf 2.3} & 82.0 & 21.0   \\
  \hline
  \multicolumn{1}{|c|}{$T_{boot}$} & 8.2 & 87.0 & 49.0   \\
  \hline
\end{tabular}
\label{tab3}
\end{center}
\end{table}

\end{document}